# Allocation d'actifs selon le critère de maximisation des fonds propres économiques en assurance non-vie : présentation et mise en œuvre dans la réglementation française et dans un référentiel de type Solvabilité 2

Frédéric PLANCHET[*]        Pierre-E. THÉROND[α]

ISFA – Université Lyon 1 [β]
WINTER & Associés [γ]

RÉSUMÉ

Le critère de maximisation des fonds propres économiques (MFPE) vise à choisir l'allocation d'actifs qui optimise, sous l'opérateur espérance, la valeur de la société d'assurance rapportée à ses fonds propres comptables. Ce critère est présenté dans le cadre d'une société d'assurance non-vie simplifiée. Il est ensuite mis en œuvre dans le cadre de la réglementation française actuelle puis d'une réglementation inspirée des travaux en cours sur le futur référentiel prudentiel européen Solvabilité 2.

Alors que dans la réglementation française, le niveau des fonds propres ne dépend pas directement de l'allocation d'actifs, il en va autrement dans un référentiel du type Solvabilité 2 puisque le capital cible doit contrôler le risque global de la compagnie auquel contribue l'allocation d'actifs. La mise en œuvre du critère de MFPE passe alors par la détermination du couple allocation d'actifs / fonds propres qui est solution d'un programme de contrôle stochastique.

Au final, une illustration numérique permet d'analyser les conséquences de la mise en place du nouveau référentiel prudentiel Solvabilité 2 sur les niveaux de provisions techniques et de fonds propres de la société puis d'illustrer l'impact du changement de référentiel sur l'allocation déterminée par le critère de MFPE. Enfin l'effet d'une mauvaise spécification de l'actif sur l'allocation déterminée par le critère de MFPE est illustré.

MOTS-CLEFS :  Allocation d'actifs, critère de maximisation des fonds propres économiques, assurance non-vie, probabilité de ruine, solvabilité 2.

*Journal of Economic Literature Classification: G11, G22 & G32.*

---

ABSTRACT

The economic equities maximization criterion (MFPE) leads to the choice of financial portfolio, which maximizes the ratio of the expected value of the insurance company on the capital. This criterion is presented in the framework of a non-life insurance company and is applied within the framework of the French legislation and in a lawful context inspired of the works in progress about the European project "Solvency 2".

In the French regulation case, the required solvency margin does not depend of the asset allocation. It is quite different in the "Solvency 2" framework because the "target capital" has to control the global risk of the company. And the financial risk takes part of this global risk. Thus the economic equities maximization criterion leads to search a couple asset allocation / equities which solves a stochastic program.

A numerical illustration makes it possible to analyze the consequences of the introduction of a "Solvency 2" framework on the technical reserves and the equities of a non-life insurance company and on the optimal allocation due to the economic equities maximization criterion. Finally, the impact of a misspecification of the risky asset model on the optimal allocation is illustrated.

*Journal of Economic Literature Classification: G11, G22 & G32.*

# 1. Introduction

Jusqu'alors dans le dispositif prudentiel français et, plus largement, dans de nombreux dispositifs européens, la solvabilité des sociétés d'assurance est assurée par une série de provisions[1] au passif et des contraintes sur les supports admissibles pour les placements à l'actif. En effet, les placements sont soumis à diverses règles prudentielles ayant pour but d'assurer leur sécurité. En particulier, leur répartition géographique et entre les différentes classes d'actifs, leur dispersion et leur congruence (cohérence de la monnaie dans laquelle ils sont libellés avec celle dans laquelle seront payées les prestations) répondent à des règles strictes[2]. Ce mode de fonctionnement a pour conséquence directe que les problématiques de détermination des fonds propres, d'une part, et d'allocation d'actifs, d'autre part, sont distinctes. En effet, dans la cadre de la réglementation européenne actuelle, le niveau minimal des fonds propres (l'exigence de marge de solvabilité) dont doit disposer un assureur dépend uniquement[3] du niveau des provisions techniques en assurance vie et du niveau des prestations et des primes en assurance non-vie.

Le projet Solvabilité 2 (*cf.* COMMISION EUROPÉENNE [2003], [2004] et AAI [2004]) en cours d'élaboration modifie profondément ces règles en introduisant comme critère explicite de détermination du niveau des fonds propres le contrôle du risque global supporté par la société. Ce risque devra notamment être quantifié au travers de la probabilité de ruine. Ainsi, la détermination de l'allocation d'actifs se trouve de fait intégrée dans la démarche de fixation du niveau des fonds propres, la structure de l'actif impactant directement la solvabilité de l'assureur.

Un certain nombre de travaux s'attachent à définir des approches standards pour la détermination du capital de solvabilité (*cf.* AAI [2004] et DJEHICHE et HÖRFELT [2005]). Dans ces approches la structure de l'actif est une donnée et les auteurs s'attachent à déterminer le niveau minimal de capital qui contrôle le risque global de la société.

Nous proposons ici un point de vue alternatif dans lequel les interactions entre le niveau du capital cible requis et l'allocation sont explicitement prises en compte et où l'on cherche à déterminer directement un couple capital / allocation d'actifs. La motivation de cette approche est qu'il nous apparaît préférable dans ce nouveau contexte de déterminer de manière conjointe le niveau du capital et la manière d'allouer ce dernier, du fait des fortes interactions que Solvabilité 2 induit à ce niveau.

Pour cela nous allons utiliser le critère de maximisation des fonds propres économiques (MFPE) initialement élaboré dans des problématiques d'assurance vie dans PLANCHET et THÉROND [2004b]. Ce critère revient à déterminer le couple capital / allocation d'actifs qui maximise la valeur initiale de la firme (exprimée en pourcentage des fonds propres réglementaires) lorsque celle-ci est mesurée sous l'opérateur espérance. Ce critère permet d'obtenir une allocation dont la détermination ne dépend pas d'un critère subjectif tel que le niveau de la probabilité de ruine. L'allocation déterminée est ensuite confrontée *ex post* à de tels indicateurs de manière à la calibrer et à valider son respect des contraintes réglementaires.

---

[1] *Cf.* art. R331-6 du Code des assurances.
[2] *Cf.* art. R332-1 et suiv. du Code des assurances.
[3] Le calcul de l'exigence de marge de solvabilité intègre également la réassurance des sociétés.

Après avoir présenté un modèle simplifié de société d'assurance sur laquelle nous allons travailler, nous reprenons la définition d'allocation optimale au sens du critère de MFPE et l'explicitons, en particulier, d'abord dans le cadre de la réglementation française puis d'un référentiel de type Solvabilité 2.

Nous illustrons ensuite la mise en œuvre de ce critère dans le cas d'une société d'assurance couvrant deux types de risques dépendants et devant composer son portefeuille financier parmi un actif risqué (dont le rendement est modélisé par un processus de Lévy simple) et un actif sans risque. Provisions techniques et niveaux de fonds propres minimaux sont déterminés dans les deux référentiels prudentiels ; puis le critère de MFPE conduit à une allocation d'actifs dans le cadre de la réglementation française et un couple capital cible / allocation d'actifs dans le référentiel de type Solvabilité 2.

Enfin l'impact d'une mauvaise spécification de l'actif sur les résultats obtenus est examiné.

## 2. Modélisation de la société d'assurance

Une société couvre sur une période[4] $n$ risques qui engendreront sur cette période les montants de sinistres $S_1, \ldots, S_n$ où $S_i$ correspond à la charge de sinistres de l'ensemble des polices de la branche $i$ dont la fonction de répartition sera notée $F_i$. Cette modélisation n'est pas restrictive dans la mesure où la variable aléatoire (v.a.) $S_i$ peut représenter les prestations effectivement versées sur la période ou leur valeur actualisée en fin de période. Nous ferons l'hypothèse que le versement des prestations a lieu en fin de période. Nous supposerons également que l'assureur ne souscrit pas de nouveau contrat en cours de période et que toutes les survenances de sinistres sont connues en fin de période.

En début de période, conformément à la réglementation, l'assureur a doté ses provisions techniques d'un montant $L_0$. Parallèlement l'assureur dispose d'un niveau de fonds propres $E_0$ qui doit être supérieur au niveau minimum de fonds propres réglementaires $E_0^R$. Nous supposerons que l'assureur place le montant $E_0 + L_0$ dans $m$ actifs financiers $A = (A_1, \ldots, A_m)$ avec les proportions[5] $\omega = (\omega_1, \ldots, \omega_m)$. Pour simplifier les notations, nous poserons, sans perte de généralité, pour tout $j \in \{1, \ldots, m\}$, $A_j = 1$ et $\sum_{j=1}^{m} \omega_j = 1$. Nous noterons $\Omega$ l'ensemble des choix de portefeuille admissibles au regard de la réglementation.

Dans le cadre de la législation française actuelle, le calcul de $L_0$ et $E_0^R$ dépend uniquement[6] de $S = (S_1, \ldots, S_n)$, alors que sous un référentiel du type Solvabilité 2, $E_0^R$ est également fonction de $\omega, A_1, \ldots, A_m$.

---

[4] Les documents de travail de la Commission européenne privilégient une approche mono-périodique pour l'appréciation de la solvabilité.
[5] On suppose donc implicitement que les règles de placement sont identiques pour les actifs en représentation des provisions techniques et pour les actifs associés aux fonds propres.
[6] Sous réserve de l'utilisation d'un principe de primes ne dépendant que des lois des $S_i$.

En fin de période, l'assureur doit payer le montant de prestations $\widehat{S} = \sum_{i=1}^{n} S_i$. Il dispose comme ressource de $(L_0 + E_0)\sum_{j=1}^{m} \omega_j A_j$. Dans la suite, nous ferons l'hypothèse que les vecteurs aléatoires $S$ et $A$ sont indépendants.

Dans la suite $\Phi$ désignera la fonction de répartition de la loi normale centrée réduite $N(0;1)$.

## 3. Critère de maximisation des fonds propres économiques

Après avoir introduit de manière générale le critère de maximisation des fonds propres économiques, nous étudions sa mise en œuvre dans le cadre de la réglementation française actuelle d'une part, puis d'un référentiel de type Solvabilité 2 d'autre part.

### 3.1. Présentation générale

En espérance, l'assureur devra débourser $\mathbf{E}[\widehat{S}] = \mathbf{E}\left[\sum_{i=1}^{n} S_i\right] = \sum_{i=1}^{n} \mathbf{E}[S_i]$ en fin de période. Par ailleurs et à la même date, ses ressources seront constituées par ses provisions techniques, ses fonds propres et les produits financiers qu'ils ont engendrés, soit $(L_0 + E_0)\sum_{j=1}^{m} \omega_j A_j$. En espérance, l'actif de l'assureur en fin de période atteindra donc le montant $\mathbf{E}\left[(L_0 + E_0)\sum_{j=1}^{m} \omega_j A_j\right] = (L_0 + E_0)\sum_{j=1}^{m} \omega_j \mathbf{E}[A_j]$.

Dans la suite nous noterons $\Lambda_0^\omega = \left(\sum_{i=1}^{n} S_i\right)\left(\sum_{j=1}^{m} \omega_j A_j\right)^{-1}$ la charge des prestations actualisée au taux de rendement du portefeuille financier et $\Sigma_0^\omega = (L_0 + E_0) - \Lambda_0^\omega$. $\Sigma_0^\omega$ peut s'interpréter comme la valeur (aléatoire) du surplus au terme actualisée au taux de rendement de l'actif.

**Définition 1 :** *Nous appellerons « fonds propres économiques » l'espérance de ce surplus actualisé* $\mathbf{E}[\Sigma_0^\omega] = \mathbf{E}[E_0 + L_0 - \Lambda_0^\omega]$ *et « provision économique » la quantité* $\mathbf{E}[\Lambda_0^\omega]$.

**Proposition 1 :** *Pour tout* $\omega \in \Omega$, *placer en début de période la quantité* $\mathbf{E}[\Lambda_0^\omega]$ *selon l'allocation* $\omega$ *permet d'être, en espérance, capable de payer les prestations en fin de période.*

***Démonstration :*** La fonction inverse étant convexe sur $]0;+\infty[$, l'inégalité de Jensen assure que pour toute variable aléatoire $X$ à valeurs strictement positives, $\mathbf{E}[X^{-1}] > (\mathbf{E}[X])^{-1}$ et donc $\mathbf{E}[\Lambda_0^\omega] * \mathbf{E}\left[\sum_{j=1}^{m} \omega_j A_j\right] \geq \mathbf{E}\left[\sum_{i=1}^{n} S_i\right]$ puisque $A$ et $S$ ont été supposés indépendants. □

Aussi les fonds propres économiques $\mathbf{E}[\Sigma_0^\omega]$ peuvent s'interpréter comme une valorisation, sous l'opérateur espérance, de la compagnie en 0. L'assureur peut rechercher à maximiser cette valorisation en référence aux capitaux de la société par le biais de son allocation d'actifs, *i. e.* à choisir le portefeuille $\omega$ qui maximise la quantité

$$\varphi(\omega) = \frac{\mathbf{E}[\Sigma_0^\omega]}{E_0}, \qquad (1)$$

*i. e.* le rapport entre la « valeur économique » de la société et ses fonds propres comptables.

**Définition 2 :** *Une allocation $\omega^*$ est optimale au sens du critère de maximisation des fonds propres économiques (MFPE) si $\omega^*$ est solution du programme d'optimisation $\sup_{\omega \in \Omega}\{\varphi(\omega)\}$.*

On note que l'ensemble des solutions $\omega^*$ de $\sup_{\omega \in \Omega}\{\varphi(\omega)\}$ coïncide avec l'ensemble des solutions du programme $\sup_{\omega \in \Omega}\{(L_0 - \mathbf{E}[\Lambda_0^\omega])/E_0\}$. Le terme $E_0^{-1}$ ne peut quant à lui être éliminé puisque l'assureur doit disposer d'un niveau de fonds propres $E_0 \geq E_0^R$ et que $E_0^R$ peut dépendre de $\omega$ (ce sera notamment le cas dans le référentiel de type Solvabilité 2 présenté *infra*).

**Proposition 2 :** *Si les variables aléatoires $\sum_{i=1}^{n} S_i$ et $\left(\sum_{j=1}^{m} \omega_j A_j\right)^{-1}$ sont indépendantes, $\omega^*$ est optimale au sens du critère MFPE si, et seulement si, $\omega^*$ est solution du programme d'optimisation $\sup_{\omega \in \Omega}\left\{E_0^{-1}\left(L_0 - \mathbf{E}\left[\sum_{i=1}^{n} S_i\right]\mathbf{E}\left[\left(\sum_{j=1}^{m} \omega_j A_j\right)^{-1}\right]\right)\right\}$.*

La démonstration de ce résultat est immédiate.

Dans la suite, on supposera que l'assureur dispose du niveau minimal de fonds propres réglementaires et donc que $E_0 = E_0^R$.

## 3.2. Mise en œuvre dans le cadre réglementaire français

La réglementation française impose d'évaluer les provisions techniques branche par branche. Elle précise[7] que ces provisions correspondent à la « valeur estimative des dépenses (…) nécessaires au règlement de tous les sinistres survenus et non payés ». Aussi le montant à provisionner en début de période au titre de la branche *i* sera l'espérance de $S_i$. Par ailleurs l'escompte des provisions est prohibé[8] de sorte que l'on a $\forall i \in \{1,...,n\}, L_0^i = \mathbf{E}[S_i]$ et $L_0 = \sum_{i=1}^{n} L_0^i$.

---

[7] *Cf.* art. R331-6 du Code des assurances.

Dans le cadre de la réglementation actuellement en vigueur, $E_0^R$ est appelé « exigence de marge de solvabilité[9] » et est indépendant de $\omega, A_1,..., A_m$. En effet, le niveau de l'exigence de marge de solvabilité est uniquement fonction des sinistres passés et des primes encaissées[10] : l'assureur calcule donc séparément l'exigence de marge selon les méthodes à partir des primes d'une part et des sinistres d'autre part, puis retient la plus grande des deux valeurs. Nous supposerons ici que l'exigence de marge de solvabilité est calculée à partir des primes. En l'absence de réassurance, la méthode à partir des primes consiste à retenir le montant $18\% \min\{\Pi_S, 50\,\text{M}€\} + 16\% \max\{\Pi_S - 50\,\text{M}€, 0\}$ où $\Pi_S$ désigne le montant total des primes commerciales encaissées. Dans la suite, nous supposerons ainsi que

$$E_0^R = 18\% * (1 + \gamma) \mathbf{E}[\hat{S}], \tag{2}$$

où $\gamma$ est le taux de chargement des primes, que nous supposerons commun à toutes les branches.

Dans ce contexte, le passif de la société n'intègre ni la dépendance qui peut exister entre les branches, ni les risques liés aux placements et, *a fortiori*, les risques actif-passif. L'indépendance de $E_0^R$ par rapport à $\omega, A_1,..., A_m$ permet d'établir trivialement la proposition suivante.

**Proposition 3 :** *Dans le cadre réglementaire français, une allocation $\omega^*$ est optimale au sens du critère de maximisation des fonds propres économiques si $\omega^*$ est solution du programme d'optimisation* $\inf_{\omega \in \Omega} \mathbf{E}\left[\left(\sum_{j=1}^{m} \omega_j A_j\right)^{-1}\right]$.

Dans ce modèle mono-périodique et en appliquant la réglementation actuelle en vigueur, l'allocation optimale au sens du critère de MFPE ne dépend donc que des caractéristiques des placements financiers. Rappelons que lorsqu'il est mis en œuvre dans un modèle multi-périodique tel que dans le cas d'un régime de rentiers (*cf.* PLANCHET et THÉROND [2004b]), la structure du passif est intégrée par le biais du profil espéré des flux futurs. Le problème d'optimisation à résoudre s'écrit alors $\inf_{\omega \in \Omega} \mathbf{E}\left[\sum_{k=1}^{\alpha}\left\{\mathbf{E}[Flux(k)]\left(\sum_{j=1}^{m} \omega_j A_j(k)\right)^{-1}\right\}\right]$, qui est une généralisation naturelle du critère présenté *supra*.

---

[8] À l'exception de branches à très long développement telles que l'assurance responsabilité civile en assurance construction par exemple.
[9] *Cf.* art. L334-1 du Code des assurances.
[10] *Cf.* Directive européenne 2002/13/CE.

Dans la situation traitée ici, le fait que le critère de MFPE ne dépende que de l'actif est donc la conséquence d'une part du fait que la marge de solvabilité ne dépend pas de ω, et d'autre part que le modèle considéré est mono-périodique.

### 3.3. Mise en œuvre dans un référentiel de type Solvabilité 2

Le projet Solvabilité 2 vise à introduire des outils de gestion de la solvabilité *globale* de la société d'assurance. En termes quantitatifs, il s'agira essentiellement :

- ✓ de déterminer, pour chaque branche, un niveau de provisions techniques qui intègre la dangerosité du risque appréciée par une mesure de risque ;

- ✓ de déterminer un niveau de « capital cible » qui contrôle, avec une forte probabilité, tous les risques supportés par la société sur un horizon fixé.

Dans la suite, nous supposerons que pour chaque branche, l'assureur doit provisionner, en début de période, le montant qui lui permet de faire face à ses engagements de payer en fin de période les prestations engendrées par cette branche dans 75% des cas. Cela revient à provisionner, pour chaque branche $i$, la Value-at-Risk[11] (VaR) à 75% du risque $S_i$ escomptée au taux d'intérêt sans risque sur la période $r$. Nous supposerons, sans perte de généralité, que la période est de durée 1.

**Définition 3 :** *La Value-at-Risk (VaR) de niveau α associée au risque X est donnée par* $\mathbf{VaR}(X,\alpha) = \inf\{x \mid \mathbf{Pr}[X \leq x] \geq \alpha\}$.

Pour chaque branche $i$, on aura donc

$$L_0^i = \mathbf{VaR}(S_i, 75\%) e^{-r}. \quad (3)$$

Par ailleurs, nous supposerons que le « capital cible » $E_0^R$ est déterminé de telle manière que l'entreprise soit capable, en fin de période, de faire face à ses engagements envers ses assurés avec une probabilité de 99,5%. Formellement cela signifie que $E_0^R$ est le plus petit montant qui remplit la condition

$$\mathbf{Pr}\left[(L_0 + E_0^R)\sum_{j=1}^{m} \omega_j A_j \geq \sum_{i=1}^{n} S_i\right] \geq 99,5\%. \quad (4)$$

Donc $E_0^R + L_0$ est la VaR à 99,5% de $\Lambda_0^\omega$. Le niveau du capital cible est donc égal à

$$E_0^R = \inf\left\{E_0 \geq 0 \mid \mathbf{Pr}\left[\Lambda_0^\omega - \sum_{i=1}^{n} \mathbf{VaR}(S_i, 75\%) e^{-r} \leq E_0\right] \geq 99,5\%\right\}. \quad (5)$$

---

[11] Nous ne discutons pas ici des différentes mesures de risque possibles ni de leur pertinence dans le cadre du projet Solvabilité 2 ; ARTZNER et al. [1999] ou encore DHAENE et al. [2004a] fournissent une présentation complète sur ce sujet.

Cette expression nous permet d'observer qu'à la différence de la marge de solvabilité, le capital cible est fonction de la dépendance stochastique entre les différentes branches et des risques liés aux placements financiers.

## 4. Application du critère de MFPE

Dans cette partie, nous allons mettre en œuvre le critère de MFPE dans le cas d'une société d'assurance qui couvre deux risques $S_1$ et $S_2$ et qui doit composer son portefeuille financier parmi deux actifs $A_1$ et $A_2$.

Après avoir comparé les bilans obtenus dans les deux systèmes prudentiels, nous comparons les allocations obtenues par le critère de MFPE et leur sensibilité à l'évolution de différents paramètres.

### 4.1. Modélisation des risques

La compagnie supporte deux types de risque : les risques de passif liés à la sinistralité engendrée par son portefeuille de contrats d'assurance et les risques de placements. Par nature ces deux types de risques sont différents aussi bien dans la manière dont ils affectent l'assureur que dans le pilotage qui peut être mis en œuvre pour les contrôler. Ainsi, à la différence des risques de sinistres, les risques de placements ne se mutualisent pas mais l'assureur peut les contrôler par sa politique d'investissement.

#### 4.1.1. Risques des sinistres

Nous supposerons que les deux risques $S_1$ et $S_2$ couverts par la société d'assurance sont distribués selon des lois log-normales $\mathbf{LN}(\mu_1, \sigma_1)$ et $\mathbf{LN}(\mu_2, \sigma_2)$. Notons $C_\alpha$ la copule de Franck qui sera supposée modéliser la dépendance entre ces deux risques. Le théorème de Sklar (*cf.* PLANCHET et al. [2005] pour une introduction à la théorie des copules) nous indique que la loi du couple $(S_1, S_2)$ peut s'écrire, de manière unique,

$$\mathbf{Pr}[S_1 \leq s_1, S_2 \leq s_2] = C_\alpha(F_1(s_1), F_2(s_2)), \quad (6)$$

où :
- $F_1$ et $F_2$, les fonctions de répartition respectivement de $S_1$ et $S_2$, sont telles que $\mathbf{Pr}[\ln S_i \leq s] = \mathbf{\Phi}\left(\dfrac{s - \mu_i}{\sigma_i}\right)$ pour $i \in \{1;2\}$,
- $C_\alpha(u_1, u_2) = -\dfrac{1}{\alpha} \ln\left\{1 + \dfrac{(e^{-\alpha u_1} - 1)(e^{-\alpha u_2} - 1)}{e^{-\alpha} - 1}\right\}$.

La copule de Franck permet de disposer d'une structure de dépendance qui permet, selon la valeur de $\alpha$, de modéliser des risques indépendants ($\alpha \to 0$), avec une dépendance négative ($\alpha \to -\infty$ conduit à la copule de la borne inférieure de Fréchet) ou avec une dépendance positive ($\alpha \to +\infty$ conduit à la copule de la borne supérieure de Fréchet). La structure de dépendance induite par cette copule est illustrée par le graphe suivant.

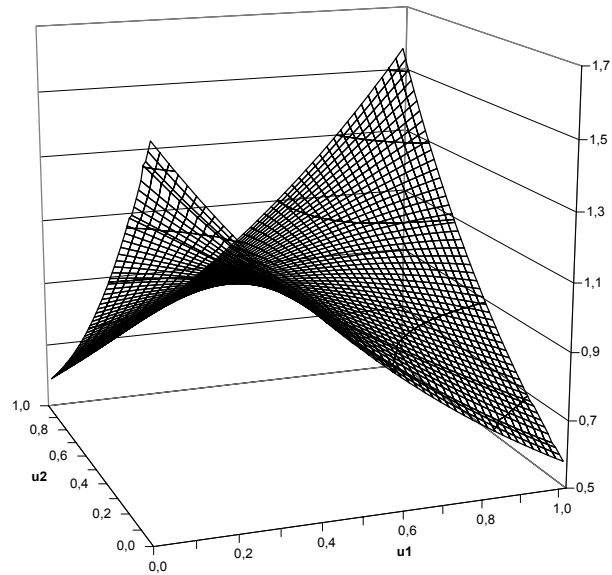

Fig. 1 - *Densité de la copule de Franck de paramètre 1*

Pour les applications numériques, nous utiliserons les paramètres suivants[12] :

$$\mu_1 = 5{,}0099 \quad \sigma_1 = 0{,}0377 \quad \alpha = 1$$

$$\mu_2 = 3{,}8421 \quad \sigma_2 = 0{,}3740 \quad \gamma = 0{,}15$$

Notons que $\alpha = 1$ correspond à des risques dont la dépendance est positive.

Comme il n'est pas possible d'expliciter par une formule directement utilisable la loi d'une somme de v.a. log-normales, nous utiliserons les techniques de Monte Carlo[13] pour obtenir la fonction de répartition empirique de cette somme. Cependant, en pratique, il apparaît souvent souhaitable de diminuer le nombre de variables à simuler, dès lors il peut être intéressant d'utiliser des approximations de leur loi. Par exemple, DHAENE et al. [2005] proposent d'approximer $\widehat{S}$ par la *comonotonic upper bound* $\widehat{S}^c$ de $\widehat{S}$ définie par

$$\widehat{S}^c = \sum_{i=1}^{2} \exp\{\mu_i + \sigma_i \Phi^{-1}(U)\}, \qquad (7)$$

où $U$ est une v.a. de loi uniforme sur $[0;1]$.

Cette approximation permet notamment de disposer de formules fermées pour calculer des Value-at-Risk ou encore des Tail-VaR. Son utilisation requiert néanmoins de mesurer l'erreur d'approximation commise. Ce point ne sera pas développé plus avant dans le présent travail.

---

[12] Les paramètres sont en fait choisis pour que l'espérance de la charge sinistre pour le risque n°1 soit égale à 150 et à 50 pour le risque n°2.
[13] En particulier, la dépendance entre les deux charges de sinistres sera intégrée à l'aide du résultat décrit en annexe.

Le graphique suivant présente la distribution de la charge totale de sinistres obtenue pour cette modélisation du passif avec les paramètres indiqués *supra*.

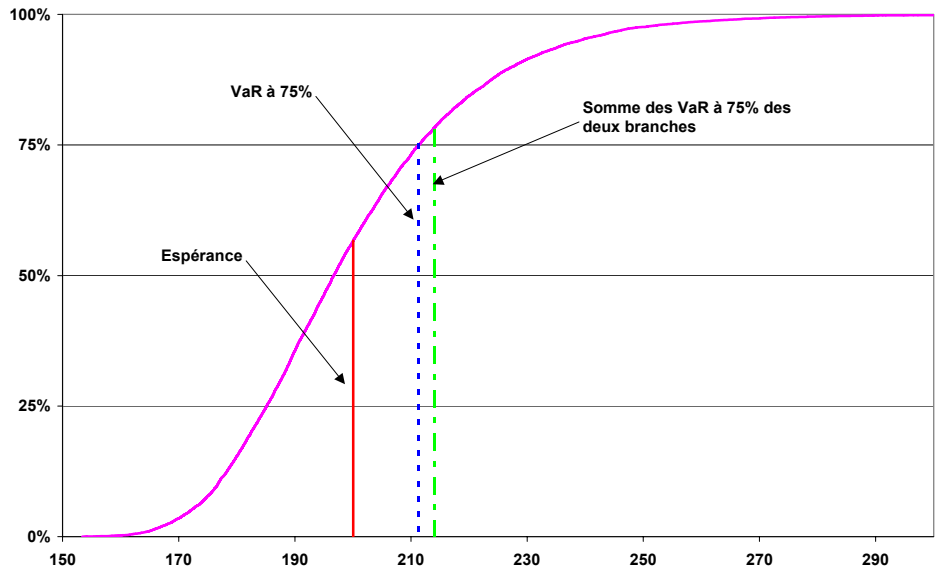

Fig. 2 - *Distribution de la charge totale de sinistres*

On remarque en particulier que la somme des VaR à 75 % est supérieure à la VaR à 75 % de la somme des risques bien que la dépendance entre les risques soit positive. On rappelle en effet que la Value-at-Risk n'est pas une mesure de risque cohérente au sens de ARTZNER et al. [1999] car elle n'est pas sous-additive.

### 4.1.2. Risques des placements

Nous supposerons que l'assureur doive composer son portefeuille financier parmi deux actifs $A_1$ et $A_2$. Pour fixer les idées, nous supposerons que $A_1$ est une actif risqué et $A_2$ un bon de capitalisation. Nous supposerons que le cours de l'actif risqué suit un processus à saut défini par

$$A_1(t) = \exp\left\{\left(\mu - \frac{\sigma^2}{2}\right)t + \sigma B_t + \sum_{k=1}^{N_t} U_k\right\}, \qquad (8)$$

où :
- $B = (B_t)_{(t \geq 0)}$ est un mouvement brownien.
- $N = (N_t)_{(t \geq 0)}$ est un processus de Poisson d'intensité $\lambda$.
- $U = (U_k)_{(k \geq 1)}$ est une suite de variables aléatoires indépendantes identiquement distribuées de loi une loi normale $\mathbf{N}(0, \sigma_u)$.
- Les processus $B$, $N$ et $U$ sont mutuellement indépendants.

Le choix de cette modélisation est motivé par la volonté de permettre de quantifier l'incidence de la présence de sauts sur la probabilité de ruine et, par-là, sur le niveau du capital de solvabilité. Aussi après avoir étudié ce processus à sauts, nous reviendrons dans le paragraphe 4.4 au cas classique du mouvement brownien géométrique. Les sauts sont ici, dans un souci de simplicité, supposés symétriques et en moyenne nuls ; des modèles plus élaborés à sauts dissymétriques peuvent également être proposés (*cf.* RAMEZANI et ZENG [1998]).

Notons $\Psi_t$ la tribu engendrée par les $B_s, N_s$ pour $s \leq t$ et $U_k \mathbf{1}_{\{k \leq N_t\}}$ pour $j \geq 1$ ; $B$ est un mouvement brownien standard par rapport à la filtration $\Psi$, $N$ est un processus adapté à cette même filtration. De plus pour tout $t > s$, $N_t - N_s$ est indépendant de la tribu $\Psi_s$.

**Proposition 4 :** *Pour tout* $x > 0$, $\Pr[A_1(t) \leq x] = \sum_{n=0}^{+\infty} \Phi\left[\dfrac{\ln x - (\mu - \sigma^2/2)t}{\sqrt{n\sigma_u^2 + t\sigma^2}}\right] e^{-\lambda t} \dfrac{(\lambda t)^n}{n!}$.

*Démonstration :* Soit $x > 0$, on a

$$\begin{aligned}
\Pr[A_1(t) \leq x] &= \Pr\left[\sum_{k=1}^{N_t} U_k + (\mu - \sigma^2/2)t + \sigma B_t \leq \ln x\right] \\
&= \sum_{n=0}^{+\infty} \Pr\left[\sum_{k=1}^{N_t} U_k + (\mu - \sigma^2/2)t + \sigma B_t \leq \ln x, N_t = n\right] \\
&= \sum_{n=0}^{+\infty} \Pr\left[\sum_{k=1}^{n} U_k + (\mu - \sigma^2/2)t + \sigma B_t \leq \ln x\right] \Pr[N_t = n],
\end{aligned}$$

puisque les processus $N$, $B$ et $U$ sont mutuellement indépendants. Par ailleurs, les processus $\sum_{k=1}^{n} U_k$ et $\sigma B_t$ étant indépendants et gaussiens, leur somme est également gaussienne : $\sum_{k=1}^{n} U_k + \sigma B_t \sim \mathbf{N}\left(0; \sqrt{n\sigma_u^2 + t\sigma^2}\right)$. Enfin comme $N$ est un processus de Poisson d'intensité $\lambda$, pour tout $t > 0$, la v. a. $N_t$ est distribuée selon une loi de Poisson de paramètre $\lambda t$ et donc $\Pr[N_t = n] = e^{-\lambda t} \dfrac{(\lambda t)^n}{n!}$. □

Lorsque $\lambda = 0$ (cas de l'absence de sauts) on retrouve la loi log-normale usuelle du brownien géométrique. Dans le cas général, l'expression de la proposition 4 permet d'approcher la distribution de l'actif en ne conservant qu'un nombre fini de termes dans la somme.

Par ailleurs, nous supposerons qu'à la date *t*, le bon de capitalisation $A_2$ vaut

$$A_2(t) = e^{rt}, \tag{9}$$

où *r* est le taux d'intérêt sans risque, supposé constant sur la période étudiée, utilisé pour escompter les provisions dans le paragraphe 3.3.

Pour les applications numériques, nous utiliserons les paramètres suivants :

$$\mu = 0{,}6 \qquad \sigma = 0{,}15 \qquad r = 0{,}344$$

$$\lambda = 0{,}5 \qquad \sigma_u = 0{,}2$$

Les sauts seront donc d'espérance nulle et, en moyenne, il en surviendra un toutes les deux périodes. Par ailleurs le taux sans risque *r* a été pris de manière à ce que le taux d'escompte discret soit de 3,5 %.

Enfin nous ferons l'hypothèse que $\Omega = [0;1] \times [0;1]$, ce qui signifie que les ventes à découvert sont interdites à l'assureur.

## 4.2. MFPE dans la réglementation française

Dans un premier temps, nous allons nous intéresser à ce qui se passe dans la réglementation française dans laquelle le niveau minimal de fonds propres ne dépend pas du choix de portefeuille et donc dans laquelle le problème d'optimisation n'intègre pas, lorsque l'on ne considère qu'une période, la variabilité des flux de sinistres.

### 4.2.1. Bilan initial

Dans le cadre de la réglementation française, l'assureur va doter ses provisions techniques en début de période du montant $L_0 = \mathbf{E}[S_1] + \mathbf{E}[S_2]$. Le montant de ses fonds propres $E_0$ sera supposé être égal à la marge de solvabilité $E_0^R$ soit : $E_0^R = 18\% * (1+\gamma)\mathbf{E}(S_1 + S_2)$. Avec les modélisations de sinistres retenues *supra*, le passif de l'assureur est donc déterminé par

$$\begin{cases} L_0 = \exp\left(\mu_1 + \frac{\sigma_1^2}{2}\right) + \exp\left(\mu_2 + \frac{\sigma_2^2}{2}\right) \\ E_0 = E_0^R = 0,18(1+\gamma)\left(\exp\left(\mu_1 + \frac{\sigma_1^2}{2}\right) + \exp\left(\mu_2 + \frac{\sigma_2^2}{2}\right)\right). \end{cases} \quad (10)$$

Avec les paramètres sélectionnés, le bilan en 0 de la société est résumé dans le tableau *infra*.

| BILAN (réglementation française) | |
|---|---|
| | $E_0 = 41{,}4$ |
| $E_0 + L_0 = 241{,}4$ | $L_0^1 = 150$ |
| | $L_0^2 = 50$ |

### 4.2.2. Allocation optimale

Dans le cadre de la réglementation française, on a vu que l'allocation $\omega = (\omega_1, 1-\omega_1)$ est optimale au sens du critère de MFPE si $\omega_1$ est solution du programme d'optimisation $\inf_{\omega \in \Omega} \mathbf{E}\left[(\omega_1 A_1 + (1-\omega_1) A_2)^{-1}\right]$.

La proposition suivante donne la condition nécessaire et suffisante pour que ce programme d'optimisation ait une solution non triviale[14].

**Proposition 5 :** *Le programme* $\inf_{\omega \in [0;1]} \mathbf{E}\left[(\omega A_1 + (1-\omega) A_2)^{-1}\right]$ *admet une unique solution* $\omega^*$ *non triviale* $(\omega^* \in \,]0;1[)$ *si, et seulement si,* $r < \mu < r + 2\sigma^2 + \lambda\left[\exp(2\sigma_u^2) - \exp(\sigma_u^2/2)\right]$.

***Démonstration :*** Cette démonstration va s'articuler en deux étapes. Dans un premier temps nous allons déterminer $\mathbf{E}\left[A_1^p(t)\right]$ pour tout réel $p$. Ce résultat nous permettra d'expliciter une condition sur les paramètres des modèles d'actifs $(r, \mu, \sigma, \sigma_u, \lambda)$ pour que $\omega^* \in \,]0;1[$.

<u>Etape 1 :</u> Soit $p \in \mathbf{R}$, on a : $\mathbf{E}\left[A_1^p(t)\right] = \mathbf{E}\left[\exp\left\{p\left(\mu - \frac{\sigma^2}{2}\right)t + p\sigma B_t + p\sum_{k=1}^{N_t} U_k\right\}\right]$. Les termes aléatoires de l'exponentielle sont indépendants ce qui ramène le calcul au produit de $\mathbf{E}\left[\exp\{p\sigma B_t\}\right]$ et $\mathbf{E}\left[\exp\left\{p\sum_{k=1}^{N_t} U_k\right\}\right]$.

Pour tout $t > 0$, $B_t$ est une v. a. de loi $\mathbf{N}\left(0; \sqrt{t}\right)$ donc

$$\mathbf{E}\left[\exp\{p\sigma B_t\}\right] = \exp\left\{\frac{p^2\sigma^2}{2}t\right\}.$$

Par ailleurs on a vu (cf. la démonstration de la proposition 4) que

$$\mathbf{E}\left[\exp\left\{p\sum_{k=1}^{N_t} U_k\right\}\right] = e^{-\lambda t} \sum_{n=0}^{+\infty} \frac{(\lambda t)^n}{n!} \mathbf{E}\left[\exp\left\{p\sum_{k=1}^{n} U_k\right\}\right].$$

Comme les sauts sont gaussiens et centrés,

$$\mathbf{E}\left[\exp\left\{p\sum_{k=1}^{n} U_k\right\}\right] = \exp\left\{\frac{np^2\sigma_u^2}{2}\right\}$$

et donc

$$\mathbf{E}\left[\exp\left\{p\sum_{k=1}^{N_t} U_k\right\}\right] = e^{-\lambda t} \sum_{n=0}^{+\infty} \frac{(\lambda t)^n}{n!} \exp\left\{\frac{np^2\sigma_u^2}{2}\right\} = \exp\left\{\lambda t\left[\exp(p^2\sigma_u^2/2) - 1\right]\right\}.$$

Ce qui permet d'obtenir

$$\mathbf{E}\left[A_1^p(t)\right] = \exp\left\{p\left(\mu - \frac{\sigma^2}{2}\right)t + \frac{p^2\sigma^2}{2}t + \lambda t\left[\exp(p^2\sigma_u^2/2) - 1\right]\right\}.$$

<u>Etape 2 :</u> La fonction objectif dans le cas réglementaire français s'écrit $\varphi(\omega) = \mathbf{E}\left[(e^r + \omega X)^{-1}\right]$ avec $X = A_1(1) - e^r$. On a donc

$$\frac{\partial \varphi}{\partial \omega}(\omega) = -\mathbf{E}\left[X(e^r + \omega X)^{-2}\right]$$

---

[14] L'expression analytique du minimum est en revanche délicate à obtenir et les calculs numériques seront menés par des techniques de simulation.

et

$$\frac{\partial^2 \varphi}{\partial \omega^2}(\omega) = 2\,\mathbf{E}\left[X^2\left(e^r + \omega X\right)^{-3}\right].$$

On en déduit trivialement que la fonction objectif est convexe : $\frac{\partial^2 \varphi}{\partial \omega^2}(\omega) > 0$. Par ailleurs

$$\frac{\partial \varphi}{\partial \omega}(0) = -e^{-2r}\,\mathbf{E}(X) = -e^{-2r}\left(e^\mu - e^r\right)$$

et

$$\frac{\partial \varphi}{\partial \omega}(1) = -\mathbf{E}\left[A_1^{-2}\left(A_1 - e^r\right)\right] = \mathbf{E}\left[A_1^{-2} e^r\right] - \mathbf{E}\left[A_1^{-1}\right].$$

Cette dernière expression se calcule à l'aide du résultat de l'étape 1 puisque

$$\mathbf{E}\left[A_1^{-2} e^r\right] = \exp\left\{r - 2\mu + 3\sigma^2 + \lambda\left(e^{2\sigma_u^2} - 1\right)\right\} \text{ et } \mathbf{E}\left[A_1^{-1}\right] = \exp\left\{-\mu + \sigma^2 + \lambda\left(e^{\sigma_u^2/2} - 1\right)\right\}.$$

Ce qui nous permet d'écrire

$$\frac{\partial \varphi}{\partial \omega}(1) = \exp\left\{r - 2\mu + 3\sigma^2 + \lambda\left(e^{2\sigma_u^2} - 1\right)\right\} - \exp\left\{-\mu + \sigma^2 + \lambda\left(e^{\sigma_u^2/2} - 1\right)\right\}.$$

La fonction $\varphi$ est strictement convexe sur $[0;1]$, elle atteint donc un unique minimum sur $]0;1[$ si $\frac{\partial \varphi}{\partial \omega}(0) < 0$ et $\frac{\partial \varphi}{\partial \omega}(1) > 1$, *i. e.* si $r < \mu < r + 2\sigma^2 + \lambda\left(\exp\left\{2\sigma_u^2\right\} - \exp\left\{\sigma_u^2/2\right\}\right)$. □

On peut vérifier que les paramètres $(r, \mu, \sigma, \sigma_u, \lambda)$ choisis en 4.1.2 sont tels qu'il existe une solution non triviale :

$$\underbrace{r}_{0,0344} < \underbrace{\mu}_{0,06} < \underbrace{r + 2\sigma^2 + \lambda\left[\exp\left(2\sigma_u^2\right) - \exp\left(\sigma_u^2/2\right)\right]}_{0,1109}.$$

Cette proposition permet d'établir que si $\mu < r$, l'assureur consacrera l'intégralité de son portefeuille financier au bon de capitalisation. En revanche si $\mu > r + 2\sigma^2 + \lambda\left(\exp\left\{2\sigma_u^2\right\} - \exp\left\{\sigma_u^2/2\right\}\right)$, provisions techniques et fonds propres seront exclusivement investis dans l'actif risqué.

Le graphique suivant présente l'évolution de la quantité $\mathbf{E}\left[\left(\omega_1 A_1 + (1 - \omega_1)A_2\right)^{-1}\right]$ en fonction de $\omega_1$.

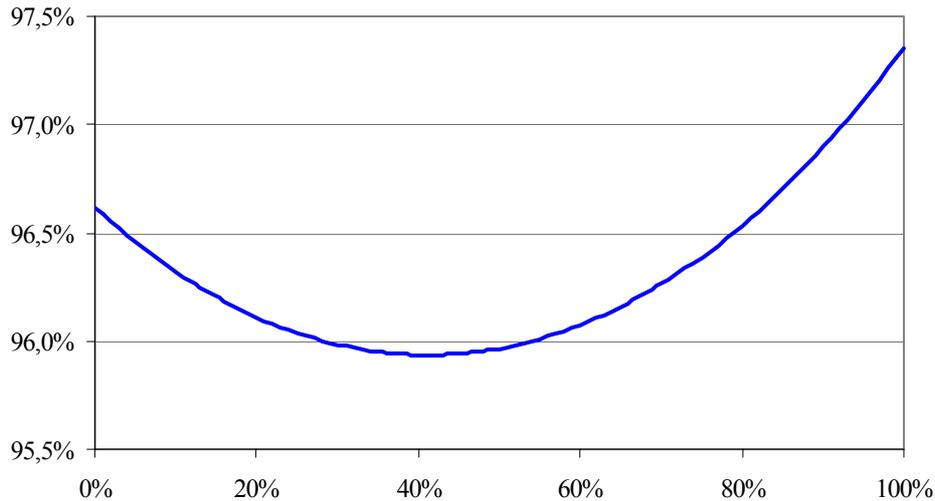

Fig. 3 - *Provision économique exprimée en pourcentage de la provision réglementaire en fonction de* $\omega_1$

Cette fonction a un minimum non trivial pour $\omega_1 = 0,391$. La société maximisera donc ses fonds propres économiques si elle place son actif initial pour 39,1 % en actions.

### 4.2.3. Probabilité de ruine

Afin de mesurer le niveau de prudence associé à l'allocation fournie par le critère de MFPE, il est naturel de déterminer la probabilité de ruine qui lui est associée. Aussi nous allons nous intéresser à la probabilité que l'assureur ne puisse payer les prestations en fin de période. Nous supposerons que la société est en faillite en fin de période si $\hat{S} > (E_0 + L_0)\left(\sum_{j=1}^{m} \omega_j A_j\right)$.

À chaque allocation $\omega$, il est possible d'associer le niveau de probabilité de ruine $\pi(\omega)$ défini par

$$\pi(\omega) = \mathbf{Pr}\left[\hat{S} > (E_0 + L_0)\sum_{j=1}^{m} \omega_j A_j\right]. \qquad (11)$$

L'expression analytique de $\pi(\omega)$ est complexe, mais cette grandeur peut être aisément approchée numériquement par des techniques de simulation[15].

Le graphique suivant reprend l'évolution de la probabilité de ruine en fonction de la part d'actifs investie en actions.

---

[15] La méthode d'obtention des réalisations de $\hat{S}$ est décrite en annexe.

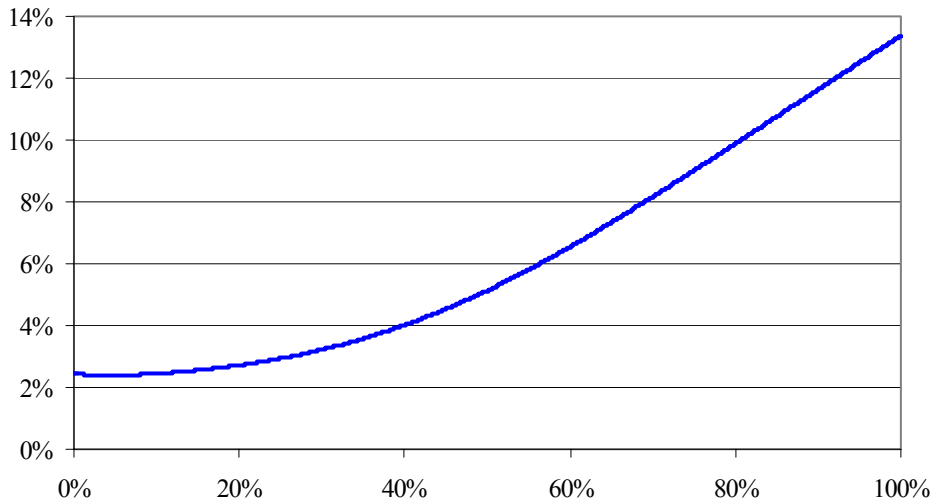

Fig. 4 - *Probabilité de ruine en fonction de* $\omega_1$

Avec un niveau de fonds propres initial égal à la marge de solvabilité, la probabilité de ruine de l'assureur en fin de période est minimale (2,4 %) lorsqu'il a placé ses provisions et ses fonds propres pour 4,3 % en actions. L'allocation optimale au sens du critère de MFPE (39,1 %) correspond quant à elle à une probabilité de ruine de l'ordre de 3,9 %.

### 4.3. MFPE dans un référentiel du type Solvabilité 2

Dans le cadre d'un référentiel de type Solvabilité 2, comme le niveau minimal de fonds propres dépend de l'allocation d'actifs, la mise en œuvre du critère MFPE passe, dans un premier temps, par la mise en lumière de la relation liant l'allocation et le capital cible ; puis par la détermination du couple capital cible / allocation qui maximise le rapport entre les fonds propres économiques et le capital cible.

#### 4.3.1. Bilan initial

Comme $S_i \sim \mathbf{LN}(\mu_i, \sigma_i)$, $\mathbf{Pr}[S_i \leq x] = \mathbf{\Phi}\left(\dfrac{\ln x - \mu_i}{\sigma_i}\right)$ et donc

$$\mathbf{VaR}(S_i, 75\%) = \exp\left\{\mu_i + \sigma_i \mathbf{\Phi}^{-1}(0,75)\right\}. \tag{12}$$

Ce qui nous permet au passage de vérifier que si les charges sinistres sont distribuées selon des lois log-normales, l'utilisation de la VaR pour déterminer le niveau des provisions est cohérente avec une approche *market value margin* qui consisterait à prendre $\exp\left\{\sigma_i \mathbf{\Phi}^{-1}(p) - \sigma_i^2/2\right\} - 1$ comme coefficient de majoration de l'espérance. En effet, la log-normalité de la charge sinistres permet d'exprimer, par le biais d'un coefficient ne dépendant que de $\sigma_i$, la VaR en fonction de l'espérance puisque

$$\mathbf{VaR}(S_i, p) = \left(1 + \exp\left\{\sigma_i \Phi^{-1}(p) - \sigma_i^2/2\right\} - 1\right) \mathbf{E}[S_i]. \tag{13}$$

Comme $L_0 = \sum_{i=1}^{2} \mathbf{VaR}(S_i, 75\%) e^{-r}$, le niveau total des provisions techniques est donné par

$$L_0 = \sum_{i=1}^{2} \exp\left\{\mu_i - r + \sigma_i \Phi^{-1}(0,75)\right\} = 148,55 + 57,97 = 206,52. \tag{14}$$

On peut noter que la faible variabilité du risque $S_1$ conduit, dans ce référentiel, à un niveau de provisions pour ce risque (148,55) inférieur à celui obtenu dans la réglementation française actuelle (150). La situation est inverse pour le risque $S_2$. Au global, le changement de référentiel prudentiel conduit à augmenter les provisions techniques de 3,26 %.

Rappelons que le niveau du capital cible $E_0^R$ dépend des risques de passif comme des risques de placement puisqu'il est solution du programme d'optimisation

$$\inf\left\{E_0 \geq 0 \mid \mathbf{Pr}\left[\Lambda_0^\omega \leq E_0 + L_0\right] \geq 99,5\%\right\}, \tag{15}$$

qui admet une unique solution car la distribution sous-jacente est absolument continue. Ce programme peut se réécrire

$$\inf\left\{E_0 \geq 0 \mid \mathbf{Pr}\left[\omega_1 A_1 + (1-\omega_1) A_2 \leq \frac{S_1 + S_2}{E_0 + L_0}\right] \leq 0,5\%\right\}. \tag{16}$$

Comme nous avons supposé que l'évolution des actifs était indépendante de la sinistralité, on a $\mathbf{Pr}\left[\omega_1 A_1 + (1-\omega_1) A_2 \leq \frac{S_1 + S_2}{E_0 + L_0}\right] = \int \mathbf{Pr}\left[\omega_1 A_1 + (1-\omega_1) A_2 \leq \frac{s}{E_0 + L_0}\right] f_S(s)\, ds$, où $f_S$ désigne la densité de la v. a. $\widehat{S} = S_1 + S_2$.

On a démontré (*cf.* proposition 4) que

$$\mathbf{Pr}\left[\omega_1 A_1 + (1-\omega_1) A_2 \leq \frac{s}{E_0 + L_0}\right] = \sum_{n=0}^{+\infty} \Phi\left[\frac{\ln \rho(s) - (\mu - \sigma^2/2)}{\sqrt{n \sigma_u^2 + \sigma^2}}\right] e^{-\lambda} \frac{\lambda^n}{n!}, \tag{17}$$

où $\rho(s) = \frac{1}{\omega_1}\left(\frac{s}{E_0 + L_0} - (1-\omega_1) e^r\right)$.

Comme $\int \sum_{n=0}^{+\infty} \Phi\left[\frac{\ln \rho(s) - (\mu - \sigma^2/2)}{\sqrt{n \sigma_u^2 + \sigma^2}}\right] e^{-\lambda} \frac{\lambda^k}{k!} f_S(s)\, ds$ n'est pas simple à calculer, nous avons choisi de résoudre numériquement ce programme d'optimisation en simulant des réalisations de $\widehat{S} = S_1 + S_2$ puis en estimant $\mathbf{Pr}\left[\Lambda_0^\omega \leq E_0 + L_0\right]$ à partir de la moyenne empirique des

$$\sum_{n=0}^{+\infty} \mathbf{\Phi}\left[\frac{\ln\rho(s_k)-(\mu-\sigma^2/2)}{\sqrt{n\sigma_u^2+\sigma^2}}\right]e^{-\lambda}\frac{\lambda^n}{n!}$$ où $s_k$ désigne une réalisation de la variable aléatoire $\widehat{S}$.

La méthode d'obtention des réalisations de $\widehat{S}$ est décrite en annexe.

En pratique nous avons développé la somme jusqu'à $n = 7$ bien que le cinquième terme de la suite soit déjà négligeable devant la somme des précédents.

Cette méthode nous permet d'obtenir la courbe suivante qui représente le niveau du capital cible en fonction de l'allocation stratégique $\omega = (\omega_1, 1-\omega_1)$.

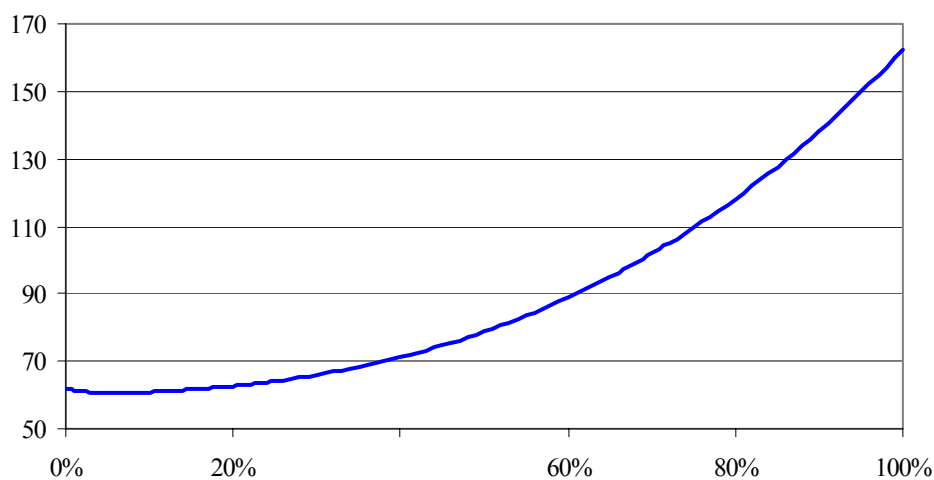

Fig. 5 - *Capital cible en fonction de la part investie en actions* $\omega_1$

Cette fonction a un minimum non trivial pour $\omega_1 = 6,1\%$ avec un capital cible de 60,71. Pour cette allocation, le passif de l'assureur s'élève à 267,24 contre 241,40 dans le cadre réglementaire français. Le passif peut atteindre 368,99 pour un actif intégralement composé d'actions.

### 4.3.2. Allocation optimale

Dans un référentiel de type Solvabilité 2, l'allocation $\omega = (\omega_1, 1-\omega_1)$ est optimale au sens du critère de MFPE si $\omega_1$ est solution du programme d'optimisation

$$\sup_{\omega_1 \in [0,1]} \{\varphi(\omega_1)\}, \tag{18}$$

où

$$\varphi(\omega_1) = \frac{\sum_{i=1}^{2}\left(\exp\{\mu_i - r + \sigma_i\,\mathbf{\Phi}^{-1}(0,75)\} - \exp\{\mu_i + \sigma_i^2/2\}\mathbf{E}\left[(\omega_1 A_1 + (1-\omega_1)e^r)^{-1}\right]\right)}{E_0^R(\omega_1)}. \tag{19}$$

Les conditions du premier ordre de ce programme sont délicates à expliciter du fait de la présence du terme $E_0^R$ au dénominateur de la fonction objectif. Aussi nous avons cherché à résoudre numériquement ce problème qui, avec les paramètres utilisés, possède une unique solution sur $]0;1[$.

Pour une allocation $\omega = (\omega_1, 1 - \omega_1)$ fixée, le montant du capital cible a été déterminé en 4.3.1 ; il reste donc à faire varier l'allocation optimale, puis à calculer pour chaque allocation la valeur de la fonction objectif.

Le graphique suivant reprend l'évolution du rapport entre les fonds propres économiques et le fonds propres réglementaires selon la part $\omega_1$ initialement investie en actif risqué $A_1$.

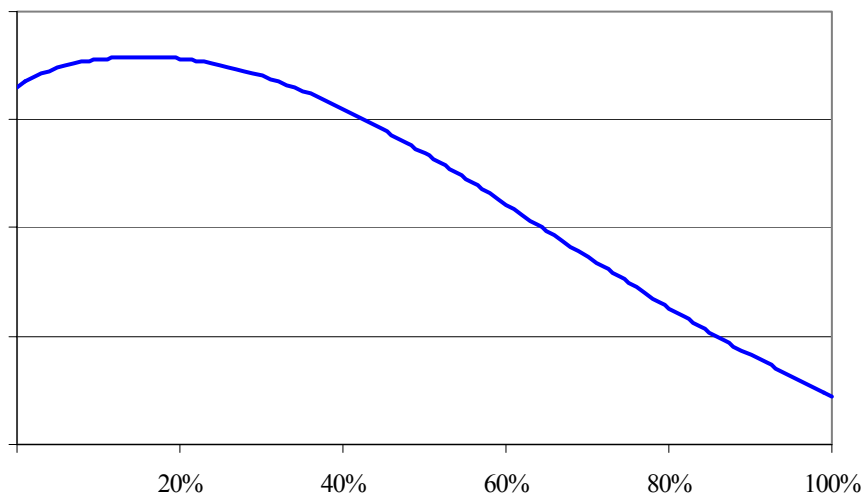

Fig. 6 - *Graphe de* φ

On constate que la fonction objectif présente un maximum sur $]0;1[$.

Si l'assureur souhaite maximiser ses fonds propres économiques, il devra composer son portefeuille financiers de 15,4 % d'actions en début de période. Pour réaliser cette allocation, les actionnaires devront fournir un capital réglementaire de 62,6. Pour cette allocation et ce capital, la société sera valorisée, sous l'opérateur espérance, à 75,7.

La probabilité de ruine définie en 4.2.2 est évidemment égale à 0,5 % puisque le capital cible a été déterminé de manière à contrôler la ruine avec cette probabilité.

## 4.4. Impact de la prise en compte des sauts de l'actif risqué

Dans un référentiel de type Solvabilité 2 dans lequel le capital cible est fonction du risque global de la compagnie, la modélisation des risques impacte directement les variables d'intérêts. L'objet de ce paragraphe est de mettre en évidence l'impact de la prise en compte des sauts de l'actif risqué par rapport au classique mouvement brownien géométrique.

Pour cela nous avons déterminé dans la situation où $\sigma_u = 0$, *i. e.* lorsqu'il n'y a pas de saut, la relation liant la part investie en actions et le niveau du capital cible.

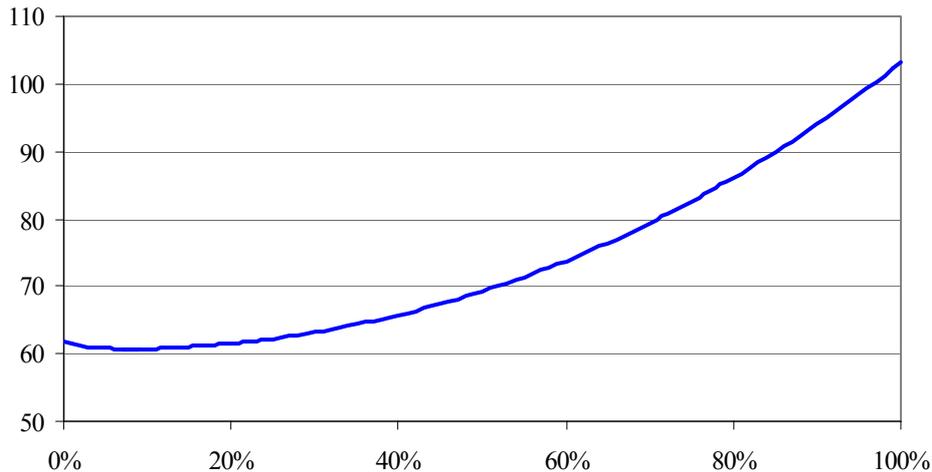

Fig. 7 - *Capital cible en fonction de la part investie en actions $\omega_1$ lorsque $\sigma_u = 0$*

Le niveau du capital cible est minimal lorsque l'actif est initialement composé de 8 % d'actions $X_1$ (contre 6,1 % précédemment).

L'évolution du capital cible en fonction de la part investie en actions permet de mesurer le risque global de la compagnie puisqu'il est déterminé de manière à contrôler la probabilité de ruine à 0,5 %. Or pour une même part investie en actions, le niveau du capital cible est systématiquement inférieur lorsque l'action suit un mouvement brownien géométrique que dans le cas où son rendement évolue selon le processus de Lévy mentionné en 4.1.2.

La graphique suivant reprend l'évolution du rapport entre le capital cible lorsque l'on prend en compte les sauts et le capital cible lorsqu'ils ne le sont pas.

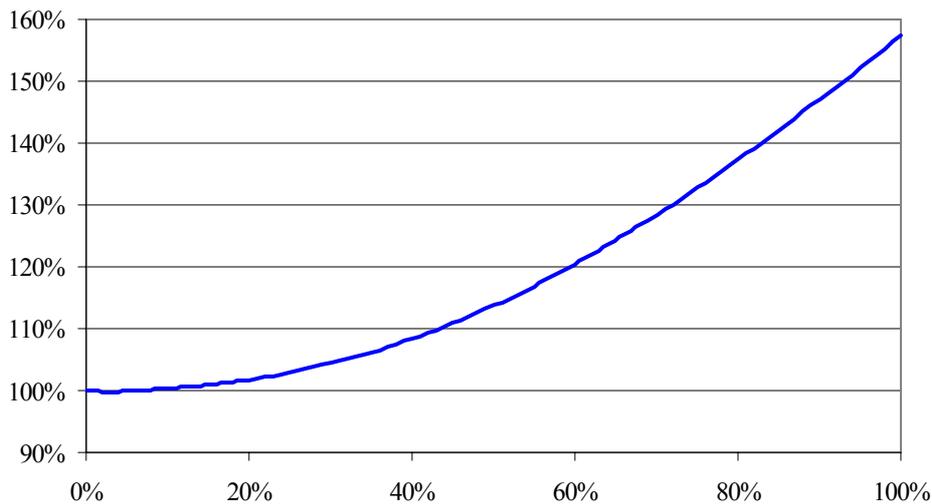

Fig. 8 - *Rapport entre le capital cible lorsque les sauts sont pris en compte et lorsqu'ils ne le sont pas*

Ce rapport est évidemment égal à 100 % lorsque le portefeuille est uniquement composé du bon de capitalisation $X_2$ et il atteint plus de 150 % lorsque l'intégralité des fonds propres et des provisions sont placés dans l'actif risqué. L'introduction d'un risque supplémentaire sur l'actif risqué a pour conséquence d'augmenter considérablement le niveau du capital cible. Ceci peut avoir une grande conséquence dans le cadre du projet Solvabilité 2 puisque comme la variabilité des actifs est difficile à mesurer, l'utilisation par deux compagnies d'assurance de modèles d'actifs renvoyant le même rendement espéré mais avec des volatilités différentes peut conduire à une distorsion des conditions de concurrence par le biais de la détermination du capital cible.

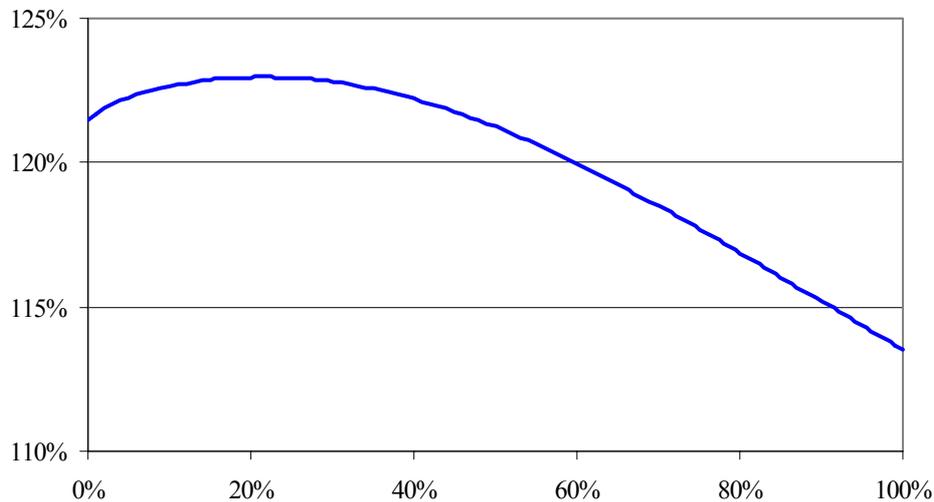

Fig. 9 - *Graphe de* φ *lorsque* $\sigma_u = 0$

Le critère de MFPE conduit à une allocation optimale de 21,4 % d'actif risqué contre 15,4 % précédemment. La prise en compte du risque supplémentaire modélisé par les sauts du rendement de l'actif risqué conduit à revoir nettement à la baisse l'allocation optimale selon le critère de MFPE.

## 5. Conclusion

Dans cet article nous avons présenté le critère de maximisation des fonds propres économiques qui consiste à composer son portefeuille financier de manière à maximiser le rapport entre l'espérance de la valeur actualisée au taux de rendement du portefeuille financier des flux futurs pour l'actionnaire et les fonds propres réglementaires.

Cette stratégie a été adaptée dans le cas d'une société d'assurance non-vie soumise au droit français puis dans celui d'une même société mais soumise à un référentiel prudentiel de type Solvabilité 2 dans lequel le niveau des fonds propres réglementaires déterminé par rapport au risque global que supporte la compagnie. Elle a été ensuite mise en œuvre pour une société couvrant deux risques dépendants et devant effectuer son choix de portefeuille dans un marché financier composé d'un actif risqué dont le rendement suit un processus de Lévy et d'un bon de capitalisation sans risque.

Dans un premier temps nous avons pu montrer l'impact sur les éléments du passif du changement de référentiel prudentiel avec, dans notre exemple, une augmentation des provisions techniques. Ensuite l'allocation optimale selon le critère de maximisation des fonds propres économiques a été déterminée.

Au final le critère proposé est séduisant car il se fonde sur un critère objectif qui ne requiert pas la détermination d'un paramètre « subjectif » tel que le niveau d'une probabilité de ruine par exemple.

Par ailleurs, nous mettons en évidence le fait que dans une approche de type « solvabilité 2 » le choix du modèle d'actif peut impacter de manière tr-s sensible le niveau du capital économique ; en particulier, les résultats obtenus donnent à penser que, dans une approche prudente, il convient d'intégrer les évolutions discontinues de l'actif, celles-ci ayant toutes choses égales par ailleurs un impact fort sur le niveau du capital. Ce point particulier fait actuellement l'objet de travaux.

## Annexe : Simulation des réalisations de la charge sinistres

Les réalisations de la charge sinistres $\hat{S}$ ont été obtenues par les techniques de Monte Carlo, à partir de la méthode des distributions conditionnelles qui permet de simuler des v. a. dont la dépendances est modélisée par une copule.

Cette méthode des distributions conditionnelles consiste à simuler[16] indépendamment deux réalisations $v_1, v_2$ de v. a. de loi uniforme sur $[0;1]$ puis d'utiliser la transformation suivante :

$$\begin{cases} u_1 = v_1 \\ u_2 = C_{u_1}^{-1}(v_2) \end{cases}$$

où $C_{u_1}(u_2) = \Pr[F_2(S_2) \leq u_2 | F_1(S_1) = u_1] = \dfrac{\partial C}{\partial u_1} C(u_1, u_2)$.

Dans le cas de la copule de Franck utilisée dans ce travail, cette dernière expression se calcule analytiquement :

$$C_{u_1}(u_2) = \frac{\exp(-\alpha u_1)(\exp(-\alpha u_2) - 1)}{\exp(-\alpha) - 1 + (\exp(-\alpha u_1) - 1)(\exp(-\alpha u_2) - 1)}.$$

De plus cette fonction s'inverse analytiquement puisque :

$$C_{u_1}^{-1}(u_2) = -\frac{1}{\alpha} \ln\left\{1 + \frac{(\exp(-\alpha) - 1) u_2}{u_2 + (1 - u_2) \exp(-\alpha u_1)}\right\}.$$

Enfin la charge totale de sinistres simulée est obtenue par

$$s = F_1^{-1}(u_1) + F_2^{-1}(u_2).$$

---

[16] Pour les illustrations numériques, les simulations de réalisations de v.a. ont été obtenues à partir du générateur du tore mélangé présenté dans PLANCHET et THÉROND [2004a].

# Bibliographie